\documentclass[%
 reprint, amsmath,amssymb, aps,showpacs]{revtex4-1}

\usepackage{graphicx}
\usepackage{epsfig}
\begin{document}


\title{Novel electric field effects on Landau levels in multi-Weyl semimetals }

\author{Amit Gupta}

\affiliation{%
Department of Physics, R. C. S. College, Manjhaul-851127, Bihar, India\\
Lalit Narayan Mithila University, Darbhanga-846008, Bihar, India  \\}%
  
\date{\today}

\begin{abstract}
Multi-Weyl semimetals(WSMs) have an anisotropic non-linear dispersion along a 2-D plane and a linear dispersion in an orthogonal direction. They have topological charge $\lq n $' and are created when two or multiple Weyl points or nodes with nonzero net monopole charge are brought together onto a high-symmetry point. We study the perturbation corrections up to second order of such multi-WSMs in crossed electric($\mathcal{E}$) and magnetic (B) fields in the low electric field approximation. As a result, the first order correction lifts only n-fold degeneracy of the lowest Landau levels(LLs) while the higher Landau levels are modified due to the second order perturbation. We study the signatures of these corrections due to electric fields on the density of states (DOS) subjected to a magnetic field.
\end{abstract}

\maketitle


\section{Introduction}
After graphene, Weyl semimetal(WSM) is another gapless system which is going to study at a rapid pace. WSM is a three dimensional analog of graphene where the low energy Hamiltonian has isotropic relativistic linear dispersion in $\textbf{k}$ space (which obey the 3D Weyl equation) from accidental degenerate band touching points referred to as Weyl points. The electronic states around the Weyl points possess a nonzero Berry curvature, which gives rise to topological charge $\pm1$ \cite{vafek14}. As a consequence, WSMs host topologically protected surface states in the form of open Fermi arcs terminating at the projections of bulk Weyl points (WPs) of opposite chirality. Other fascinating physical consequences of these Berry phases are exotic transport phenomena such as a large negative magnetoresistance due to chiral anomaly \cite{volovik86,volovik09,son13}. In recent angle-resolved photoemission spectroscopy(ARPES) and scanning tunneling microscopy(STM) experiments, several materials such as Cd$_3$As$_2$ \cite{neupane14,jeon14,liu14,borisenko14,he14,li15,moll15}, Na$_3$Bi \cite{liu14a},NbAs\cite{xu15a}, TaP\cite{xu15,xu16}, ZrTe$_5$ \cite{chen15,li16} and TaAs \cite{lv15,lv15a,lv15b,huang15a,xu15b,yang15,inoue16} have been identified as Weyl semimetals. Also, several attempts have been made on the realization of WSMs in artificial systems such as photonic crystals \cite{chen15a,zhang15,dubcek15,lu15,lu16}.

In addition to above  isotropic or single WSM, a new three-dimensional topological semimetals have been proposed in materials with certain point-group characterized by $C_ {4,6}$ symmetries \cite{xu11,fang12,huang16} e.g. the double(triple)-Weyl semimetals have band touching points with quadratic(cubic) dispersions along $k_x-k_y$ plane and linear dispersion along $k_z$ direction. The double-Weyl nodes are protected by $C_4$ or $C_6$ rotation symmetry and its half-metallicity has been realized in the three-dimensional semimetal $\mathrm{HgCr_2Se_4}$ in the ferromagnetic phase, with a  single pair of double-Weyl nodes along the z-direction \cite{guan15}. However, the material realization of triple-WSMs remains elusive. The double-Weyl node possesses a monopole(anti-monopole) charge of +2 (-2) and it shows double-Fermi arcs on the surface Brillouin zone(BZ) \cite{xu11,fang12,huang16}. Similarly, triple-Weyl node possesses a monopole(anti-monopole) charge of +3 (-3) and it shows triple-Fermi arcs on the surface BZ.\\

The single or isotropic-WSMs are predicted to have a fascinating response similar to graphene and 2D- WSM in crossed electric ($\mathcal{E}$) and magnetic fields (B) \cite{ming16}. These systems have topologically protected gapless Dirac or Weyl nodes with relativistic dispersion. This naturally induces Lorentz boosts in crossed electric ($\mathcal{E}$) and magnetic fields (B) \cite{luckose07}. As a consequence of Lorentz invariance of the Dirac equation, these problems can be solved exactly by choosing a reference frame in which the electric field vanishes as long as the drift velocity $v_d=\mathcal{E}/\mathrm{B} $ is smaller than the Fermi velocity($v_F$), which plays the role of an upper bound for the velocity as the speed of light c in relativity. This lifts the Landau levels(LLs) degeneracy. As a consequence, the LLs spacing is reduced and at a particular value of $\mathcal{E}/ v_F \mathrm{B}$, the LLs get collapse \cite{ming16,luckose07}. \\

Our main contribution is to extend this mechanism for multi-WSMs(double and triple-WSMs). The double and triple-WSMs have non-Lorentz invariant physics due to quadratic and cubic dispersions respectively in $k_x-k_y$ plane and therefore their crossed electric and magnetic fields response cannot be solved exactly due to the absence of a reference frame in which the electric field vanishes. We, therefore, study this problem by perturbation theory up to second order corrections in electric fields as it has been discussed for multilayers graphene \cite{katsnelson13}. The rest of the paper is organized as follows. In Section \ref{sec:2}, we study the Landau level spectrum of the multi-Weyl semimetals in the presence of an in-plane uniform electric field. These problems have been well studied in 2-dimensional single layer graphene, 2-D WSMs and 3-dimensional single WSMs and type-II WSMs\cite{luckose07,peres07,goerbig09,morinari09,tchoumakov16}. In Section \ref{sec:3}, we study the response of multi-WSMs in crossed electric and magnetic fields by perturbation theory. We also calculate the density of states in crossed fields in Section \ref{sec:4}. Finally, we make some concluding remarks in Section \ref{sec:5}.

\section{Landau levels formation in multi-Weyl semimetals}
\label{sec:2}
The non-interacting low energy effective Hamiltonian for a single multi-Weyl semimetals is given by \cite{ahn16, bitan15, lia16, ahna16},
\begin{equation}
\label{eq:ham}
H=\alpha_n[( \hat{p}_{-})^n \sigma_{+}+ (\hat{p}_{+})^n\sigma _{-}]+\chi v_z\hat{p}_z\sigma _z,
\end{equation}
where $\sigma_{\pm}=\frac{1}{2}(\sigma_{x}\pm i\sigma_{y})$, $\hat{p}_{\pm}=\hat{p}_x\pm i \hat{p}_{y}$,  and $\chi =\pm 1$ is the chilarity, $\lq n $' represents monopole charge , $v_z$ is the Fermi velocity along $\hat{z}$ direction and  $\alpha_{n}$ is the material dependent parameter, e.g. $\alpha_1$ and $\alpha_2$ are the Fermi velocity and inverse of the mass respectively for single and double WSMs. The energy spectrum of Eq.(\ref{eq:ham}),
\begin{equation}
\epsilon _{s}(\mathbf{k})=s\sqrt{\alpha_n^2 (\hbar k_{\parallel})^{2n}+\left(\hbar k_z v_z \right)^2},
\end{equation}

\noindent where $s=\pm 1$ and $ k_{\parallel}=\sqrt{k_x^2+k_y^2} $ is the momentum along $\hat{x}$-$\hat{y}$ plane. The density of states of such an anisotropic Hamiltonian is given by $g(\epsilon )\sim \epsilon^{2/n}$ \cite{ahn16}.  In the presence of large external magnetic field, the Hamiltonian Eq.(\ref{eq:ham}) form the Landau level spectrum \cite{Abrikosov98,Roy15,Chen16a}. Under an external magnetic field $ \textbf{B} $ directed along $z$-axis, we make the usual Peierls substitution $ \textbf{p}\rightarrow \textbf{p}+e\textbf{A} $ ($e>0$) in Hamiltonian Eq. (\ref{eq:ham}) with the vector potential $\textbf{A}$. We choose $ \textbf{A} $ in the Landau gauge  ($A_{y}=Bx$). Therefore, the Hamiltonian Eqn.$(1)$ transforms to

\begin{equation}
H=\left(
\begin{array}{cc}
v_z\hat{p}_z & \alpha_n\Bigl(\hat{p}_{x}-i(\hat{p}_{y}+eBx)\Bigr)^{n} \\
\alpha_n\Bigl(\hat{p}_{x}+i(\hat{p}_{y}+eBx)\Bigr)^{n} & -v_z\hat{p}_z
\end{array}
\right)   \label{eq1}
\end{equation}%

\noindent where $\hat{p}_{x, y, z}=-i\hbar \partial _{x, y, z}$. The time-independent Schr\"odinger equation $H\Psi = E_m\Psi$. Since the above Hamiltonian contain explicitly only x, we will look for the solutions of the usual form
\begin{equation}
\Psi=\Psi(x)e^{i\hbar(k_yy+k_zz)}
\end{equation}

so that Eqn.$(3)$ transforms to

\begin{widetext}
\begin{eqnarray}
\left(
\begin{array}{cc}
\label{Big:eqn}
\hbar v_zk_z & \alpha_n \Bigl(\hat{p}_{x}-i(\hbar{k}_{y}+ eBx)\Bigr)^{n} \\
\alpha_n \Bigl(\hat{p}_{x}+i(\hbar{k}_{y}+ eBx)\Bigr)^{n} & -\hbar v_zk_z%
\end{array}\right) \Psi(x)=E\Psi(x)
\end{eqnarray}
\end{widetext}

Let us now define a new variable $ \quad u=\frac{(x+ k_{y}l_B^2)}{l_B}$ with $l_B=\sqrt{\hbar/(eB)}$ the magnetic length and introduce the annihilation and creation operators $\hat{a}=\frac{1}{\sqrt{2}}[u+\partial _{u}],\hat{a}^{\dag }=\frac{1}{\sqrt{2}%
}[u-\partial _{u}]$ which satisfy the standard commutation relation $ [\hat{a},\hat{a}^{\dag }]=1 $. Thus, Eqn.$(\ref{Big:eqn})$ becomes

\begin{eqnarray}
\left(
\begin{array}{cc}
\hbar v_zk_z & (-i)^n\alpha_n(\sqrt{2}\frac{\hbar}{l_B})^n\hat{a}^n  \\
(i)^n\alpha_n(\sqrt{2}\frac{\hbar}{l_B})^n[\hat{a}^{\dag }]^n & -\hbar v_zk_z
\end{array}\right) \Psi_m = \nonumber\\
E_m \Psi_m\nonumber\\
\end{eqnarray}

with Landau levels(LLs) spectrum for $ m\geq n$
\begin{eqnarray}
 E_m ^0&=&s\sqrt{ m(m-1)...(m-n+1)(\omega)^{2n}\alpha_n^2 +v_z^2\left(\hbar k_z\right)^2} \nonumber\\
\end{eqnarray}

whereas for $m<n$

\begin{equation}
E_{0,1,2,...n}=\chi \hbar v_zk_z  \label{eq_lls}
\end{equation}

\noindent with $ \omega=\frac{\sqrt{2}\hbar}{l_B}$ and  m is the Landau level index. Thus, it is clear from Eq.(\ref{eq_lls}) that the lowest Landau levels(LLS) have n- degenerate chiral modes i.e. the lowest LLs for double and triple WSM have \emph{two} and \emph{three} fold degeneracy, respectively.

The normalized solutions for  $m\geq n$ are

\begin{equation}
\Psi _{m, s=+}=\frac{1}{\sqrt{2}}\left(
\begin{array}{c}
(-i)^n b_m\psi_{m-n}\\
a_m \psi_{m}
\end{array}
\right)
\end{equation}

\begin{equation}
\Psi _{m, s=-1}=\frac{1}{\sqrt{2}}\left(
\begin{array}{c}
-(-i)^n a_m\psi_{m-n}\\
b_m \psi_{m}
\end{array}
\right)
\end{equation}

whereas for $m<n$
\begin{equation}
\Psi _{n}=\left(
\begin{array}{c}
0\\
\psi_{n} 
\end{array}
\right)
\end{equation}

\noindent where $a_m =\Bigl(1+\frac{v_z\hbar k_z}{E_m^0}\Bigr)^{1/2}$, $b_m =\Bigl(1-\frac{v_z\hbar k_z}{E_m^0}\Bigr)^{1/2}$ and $\psi_m $ are the usual normalized eigenfunctions of a free electron in a magnetic field
\begin{equation}
\psi_m=\frac{\sqrt{l_B}}{\sqrt{2^m m!\sqrt{\pi }}}e^{-u^{2}/2}H_{m}(u) 
\end{equation}

Here $ H_m $ are Hermite polynomials. 


\section{Energy levels in the presence of magnetic field and electric field}
\label{sec:3}
Let us assume that in addition to the magnetic field, one has an uniform electric field along the x direction. This add to the Hamiltonian a term of the form $-e V\textbf{1}=-e\mathcal{E}x\textbf{1}$ where V is the electric potential associated with the applied electric field $ \textbf{E}=(\mathcal{E},0,0) $ and $ \textbf{1} $ is the $ 2 \times 2$ unit matrix. Then the single particle Hamiltonian is given by 

\begin{equation}
H=\left(
\begin{array}{cc}
v_z\hat{p}_z-e\mathcal{E}x & \alpha_n\Bigl(\hat{p}_{x}-i(\hat{p}_{y}+eBx)\Bigr)^{n} \\
\alpha_n\Bigl(\hat{p}_{x}+i(\hat{p}_{y}+eBx)\Bigr)^{n} & -v_z\hat{p}_z-e\mathcal{E}x
\end{array}%
\right)   \label{eqeb}
\end{equation}

We try the wave function as $\Psi_{1,2}=\Psi_{1,2}(x)e^{i\hbar(k_yy+k_zz)}$. For n=1, the above $2\times 2$ Hamiltonian is an equivalent to a tilted WSMs \cite{ming16} and can be exactly solved for $\mathcal{E}/v_FB<1$ due to Lorentz invariant physics of the Hamiltonian. The corresponding eigenvalues are given by

\begin{equation}
E_m=\frac{1}{\gamma}\sqrt{v_z^2\hbar^2 k_z^2 + \frac{2eB\hbar}{\gamma}m \alpha_1^2}+\frac{\mathcal{E}}{B}\hbar k_y 
\end{equation}

\noindent where $\gamma=1/\sqrt{1-\beta^2}$, $\beta=\mathcal{E}/{\alpha_1B} $. Therefore, when $\mathcal{E}/{\alpha_1B}=1$, there is a complete collapse of the Landau Levels(LLs).
The above Hamiltonian is not exactly soluble for topological charge $n >1$  due to the absence of a reference frame in which the electric field vanishes and therefore we solve the above problem for low electric field and high magnetic field such that Landau levels remain intact.

\subsection{The zero order approximation}

Let us introduce an another parameter $ \lambda $
\begin{equation}
\lambda = \frac{e \mathcal{E}l_B}{\sqrt{2}} 
\end{equation}

\noindent such that $ \lambda$ is always small with respect to the leading energy scale $2^{n/2}\hbar^n \alpha_n/l_B^n$ which holds at low electric field $ \mathcal{E}$ . This  validates the perturbation theory. Now, the Schr\"{o}dinger equation (\ref{eqeb}) reads

\begin{widetext}
\begin{eqnarray}
\left(
\begin{array}{cc}
\hbar v_zk_z -\lambda (\hat{a}^{\dag }+\hat{a}) & (-i)^n 2^{n/2}\frac{\hbar^n}{l_B^n}\alpha_n \hat{a}^n  \\
 (i)^n 2^{n/2}\frac{\hbar^n}{l_B^n}\alpha_n[\hat{a}^{\dag }]^n & -\hbar v_zk_z -\lambda (\hat{a}^{\dag }+\hat{a})
\end{array}\right) \Psi_{1,2}(x)&=&\tilde{E}_m \Psi_{1,2}(x)= (E_m-\frac{\mathcal{E}}{B}\hbar k_y) \Psi_{1,2}(x)\nonumber\\
\end{eqnarray}
\end{widetext}

When $\mathcal{E}$ is small, in the zeroth order approximation over $\lambda$ we have:
\begin{eqnarray}
\hbar v_zk_z\psi_1 + (-i)^n 2^{n/2}\frac{\hbar^n}{l_B^n}\alpha_n[\hat{a}]^{n}\psi_2 &=&(E_m-\frac{\mathcal{E}}{B}\hbar k_y){\psi }_{1}  \nonumber \\
(i)^n 2^{n/2}\frac{\hbar^n}{l_B^n}\alpha_n[\hat{a}^{\dag }]^{n}\psi _{1}-\hbar v_zk_z\psi_2 &=&(E_m-\frac{\mathcal{E}}{B}\hbar k_y){\psi }_{2}\nonumber\\
\end{eqnarray}
(We cannot neglect here $ \frac{\mathcal{E}}{B}\hbar k_y$ because $\hbar k_y$ may be large.) Thus, we have the same Landau levels as without electric field just shifted by $ \frac{\mathcal{E}}{B}\hbar k_y$.\\

\begin{figure}[h]       
\fbox{\includegraphics[scale=.40]{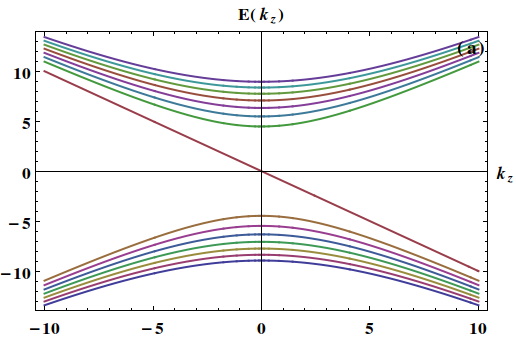}}   
\hspace{30px}
\fbox{\includegraphics[scale=.3]{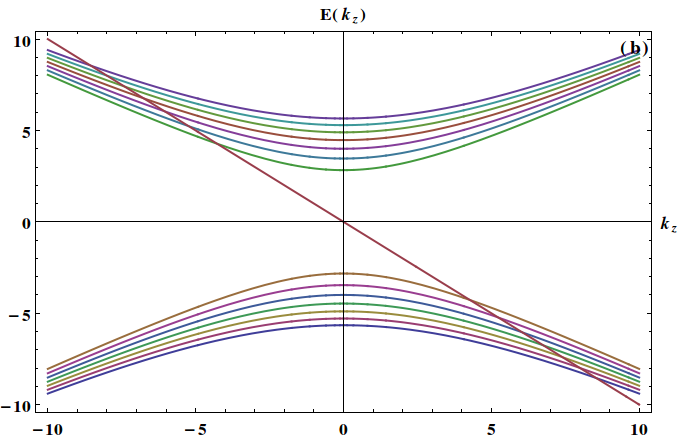}}
\caption{Landau levels spectrum for a single-Weyl semimetals for $k_y=0$ (a)obtained exactly for $\mathcal{E}=0$ and (b) obtained in perturbation theory to the second order in electric field for  $\mathcal{E}=3.5$. Other parameters: B=5, $v_z$ =$1$, $\alpha_2=1$.}
\label{figswsm}
\end{figure}

\subsection{The first order correction}

In the first order approximation, we have

\begin{widetext}
\begin{eqnarray}
\left(
\begin{array}{cc}
\hbar v_zk_z -\lambda (\hat{a}^{\dag }+\hat{a}) & (-i)^n 2^{n/2}\frac{\hbar^n}{l_B^n}\alpha_n \hat{a}^n  \\
 (i)^n 2^{n/2}\frac{\hbar^n}{l_B^n}\alpha_n[\hat{a}^{\dag }]^n & -\hbar v_zk_z -\lambda (\hat{a}^{\dag }+\hat{a})
\end{array}\right) \Psi_{1,2}(x)=\tilde{E}_m \Psi_{1,2}(x) = (E_m-\frac{\mathcal{E}}{B}\hbar k_y) \Psi_{1,2}(x)
\end{eqnarray}
\end{widetext}

One can easily see that the first-order term due to $\lambda $ vanishes for $m\ge n$. For $m<n$, we use degenerate first order perturbation theory \cite{capri} and finds that there is the first order correction to the energy for $n\ge 2$. This correction is given by 

\begin{equation}
\Phi_{m}^{\dagger}H'\Phi_{m'}=\tilde{E}_{m}^{(1)}\delta_{mm'}
{\label{degenerate_first}}
\end{equation}

\noindent where $H'=-\lambda (\hat{a}^{\dag }+\hat{a}) \mathbb{I}_{2 \times 2}$ and $\Phi_{m}$ is a new basis which has been expressed in terms of original lowest Landau levels basis $\Psi_m$ by unitary transformation

\begin{equation}
\Phi_{m}=\sum_{m'=0}^{n-1}a_{m'}\Psi_{m'}
\end{equation}
where $a_{m'}$ is the ampitude of $\Psi_{m'}$ and summation is over lowest Landau levels  (i.e.$m'<n$). The matrices $\Omega$ formed from $\Psi_p'H'\Psi_m'$ for double and triple-WSMs reduce to

\begin{equation}
\left(\begin{array}{cc}0&-\lambda\\
-\lambda & 0 
 \end{array}\right)=\Omega
\end{equation}

and 
\begin{equation}
\left(\begin{array}{ccc}0&-\lambda &0\\
-\lambda & 0 & -\sqrt{2}\lambda\\
0 & -\sqrt{2}\lambda & 0
 \end{array}\right)=\Omega
\end{equation}

respectively which have eigenvalues
\begin{eqnarray}
&&\tilde{E}_{m}^{(1)} = \pm \lambda;\nonumber\\
&&\tilde{E}_{m}^{(1)} = 0, \pm \sqrt{3}\lambda 
\end{eqnarray}
for double and triple-WSMs respectively. Therefore, in the presence of electric field $\mathcal{E}$, the n-fold degeneracy of chiral lowest LLs is lifted.\\

\subsection{The second order approximation}

We have already seen that the first-order term in $\mathcal{E} $ in powers of $\lambda $ vanishes for $m\ge n$. Therefore, we look at the second-order term

\begin{equation}
\tilde{E}_{m,s}^{(2)}=\lambda ^{2}\sum_{k\neq m,s'}\frac{1}{E _{m,s}^{(0)}-E _{k,s'}^{(0)}}|\Psi_{m,s}^{\dagger}(\hat{a}^{\dag }+\hat{a})\Psi_{k,s'}|^{2}
\label{ute_2} 
\end{equation}

\noindent where s, $s'$ denotes the band index. \\\\

First, we consider the case $m\geq n$  for which Eq. (\ref{ute_2}) can be rewritten as

\begin{eqnarray}
\tilde{E}_{m}^{(2)} &=&\frac{\lambda ^{2}
}{4}\Bigl(\frac{1}{E _{m}^{(0)}-E_{m+1}^{(0)}}[a_{m+1}a_m\sqrt{m+1}+b_{m+1}b_m\sqrt{m-n+1}]^{2}\nonumber \\
&&+\frac{1}{E_{m}^{(0)}-E_{m-1}^{(0)}}[a_ma_{m-1}\sqrt{m}+b_mb_{m-1}\sqrt{m-n}]^{2}
\nonumber\\  
&&+\frac{1}{E_{m}^{(0)}+E_{m+1}^{(0)}}[a_mb_{m+1}\sqrt{m+1}-a_{n+1}b_n\sqrt{m-n+1}]^{2}
\nonumber\\
&&+\frac{1}{E_{m}^{(0)}+E_{m-1}^{(0)}}[a_mb_{m-1}\sqrt{m}-a_{m-1}b_{m}\sqrt{m-n}]^{2}\Bigr)\nonumber\\
&&=\frac{\lambda^2}{4}\zeta_m \label{ute_2_}
\end{eqnarray}

with

\begin{eqnarray}
\zeta_m &=&\frac{2}{\theta_1}\Bigl[\{2(m+1)-n\}E_m^{0}+2n\hbar v_z k_z+\{2(m+1)-n\}\frac{(\hbar v_z k_z)^2}{E_m^0}\nonumber\\
&&+\frac{2}{E_m^0}(m+1)m(m-1)....(m-n+1)\omega^{2n}\alpha_n^2\Bigr] \nonumber\\
&&\frac{2}{\theta_2}\Bigl[(2m-n)E_m^{0}+2n\hbar v_z k_z+(2m-n)\frac{(\hbar v_z k_z)^2}{E_m^0}\nonumber\\
&&+\frac{2}{E_m^0}m(m-1)....(m-n)\omega^{2n}\alpha_n^2\Bigr] 
\end{eqnarray}

\noindent where $ \theta_1=(E_{m}^{0})^2-(E_{m+1}^{0})^2=-n m(m-1)(m-2)....(m-n+2)\omega^{2n}\alpha_n^2 $ and $ \theta_2=(E_{m}^{0})^2-(E_{m-1}^{0})^2=n(m-1)(m-2)....(m-n+2)(m-n+1)\omega^{2n}\alpha_n^2 $.\\

In particular, for a single WSMs case ($n=1$), the second order energy correction
\begin{eqnarray}
\tilde{E}_{m}^{(2)}&=&-\frac{1}{4}\beta^2[2E_m^0+\frac{m}{E_m^{0}}(\alpha_1^2 \omega^2)]\nonumber\\
&&=-\frac{1}{2}\beta^2 E_m^{0}-\frac{1}{2}\frac{m}{E_m^{0}}(\hbar eB)\alpha_1^2 \beta^2
\end{eqnarray}

\begin{figure}[h]       
\fbox{\includegraphics[scale=.30]{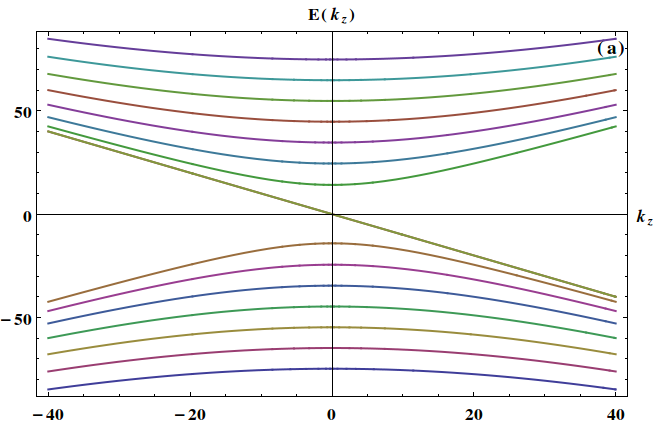}}   
\hspace{30px}
\fbox{\includegraphics[scale=.35]{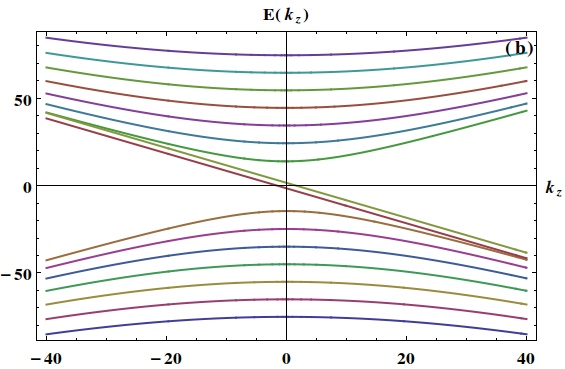}}
\caption{Landau levels spectrum for double-Weyl semimetals for $k_y=0$ (a) obtained exactly for $\mathcal{E}=0$ and (b) obtained in perturbation theory to the second order in the electric field for  $\mathcal{E}=5$. Other parameters: B=5, $v_z$ =$1$, $\alpha_2=1$.}
\label{figdwsm}
\end{figure}

\begin{figure}[h]       
\fbox{\includegraphics[scale=.34]{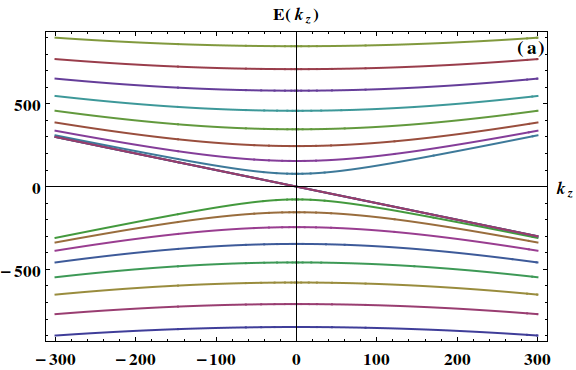}}   
\hspace{30px}
\fbox{\includegraphics[scale=.35]{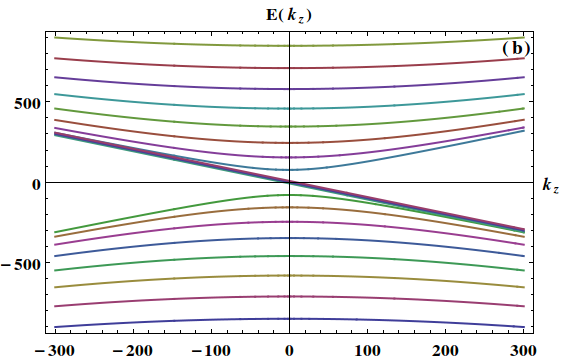}}
\caption{Landau levels spectrum for triple-Weyl semimetals for $k_y=0$ (a)obtained exactly for $\mathcal{E}=0$ and (b)obtained in perturbation theory to the second order in the lectric field for $\mathcal{E}=25$. Other parameters: B=5, $v_z$ =$1$, $\alpha_2=1$.}
\label{figtwsm}
\end{figure}

Therefore, the modified energy

\begin{eqnarray}
E_m&=&E_m^0-\frac{1}{2}\beta^2 E_m^0-\frac{1}{2}\frac{m}{E_m^{0}}(\hbar eB)\alpha_1^2 \beta^2+\frac{\mathcal{E}}{B}\hbar k_y \nonumber\\
&&\approx\frac{1}{\gamma}E_m^0-\frac{1}{2}\frac{m}{E_m^{0}}(\hbar eB)\alpha_1^2 \beta^2+\frac{\mathcal{E}}{B}\hbar k_y 
\end{eqnarray}
where we have taken $\gamma=1/\sqrt{1-\beta^2}\approx 1/(1-\frac{1}{2}\beta^2)$ at low $\beta$ in above equation. This energy agrees with exact results for type-II WSMs or tilted single WSMs in ref.\cite{tchoumakov16}.

\begin{eqnarray}
E_m(k_z)=\frac{1}{\gamma}\sqrt{v_z^2\hbar^2 k_z^2 + \frac{2eB\hbar}{\gamma}m \alpha_1^2}+\frac{\mathcal{E}}{B}\hbar k_y
\end{eqnarray}
 
\noindent Similarly, we can show that Landau levels spectrum of other multi WSMs gets modified in the presence of electric field. For double-WSMs, the second order energy correction\\

\begin{eqnarray}
\tilde{E}_{m}^{(2)}&=&\lambda^2 \Bigl[-\frac{(2m-1)}{2 E_m^{0}}+\frac{\hbar v_z k_z}{\omega^4 \alpha_2^2 m(m-1)}\Bigr]
\end{eqnarray}
 and for triple-WSMs,we have second order energy correction
\begin{eqnarray}
\tilde{E}_{m}^{(2)}&=&\lambda^2 \Bigl[ \frac{(E_m^{0})^2+(\hbar v_z k_z)^2}{3E_m^{0} \omega^6 \alpha_3^2 m(m-2)}-\frac{2(m-1)}{3 E_m^{0}}\nonumber\\
&&+\frac{2\hbar v_z k_z}{m(m-1)(m-2)\omega^6 \alpha_3^2}\Bigr]
\end{eqnarray} 

Therefore, the corresponding modified energies for double WSMs is
 
\begin{eqnarray}
E_m&=&E_m^0+\lambda^2 \Bigl[-\frac{(2m-1)}{2 E_m^{0}}+\frac{\hbar v_z k_z}{\omega^4 \alpha_2^2 m(m-1)}\Bigr]+\frac{\mathcal{E}}{B}\hbar k_y\nonumber\\
\label{eqn_dwsms}
\end{eqnarray}

\noindent and for triple-WSMs, the modified energy dispersion

\begin{eqnarray}
E_m&=&E_m^0+\lambda^2 \Bigl[ \frac{(E_m^{0})^2+(\hbar v_z k_z)^2}{3E_m^{0} \omega^6 \alpha_3^2 m(m-2)}-\frac{2(m-1)}{3 E_m^{0}}\nonumber\\
&&+\frac{2\hbar v_z k_z}{m(m-1)(m-2)\omega^6 \alpha_3^2}\Bigr]+\frac{\mathcal{E}}{B}\hbar k_y
\label{eqn_twsms}
\end{eqnarray}

For m$<$n, we use degenerate perturbation theory(Rayleigh-Schr\"{o}dinger solution)\cite{capri} and and show it easily that the second order correction vanishes due to the cancellation from positive and negative energy bands of higher Landau levels. Therefore, the modified lowest LLs energy dispersion due to the first and second order corrections.

\begin{equation}
E_m=-\hbar k_z v_z +\frac{\mathcal{E}}{B}\hbar k_y -\lambda\mathrm{EV}[\Omega]
\end{equation}
\label{eqn_llls}
where EV$(\Omega)$ represents the eigenvalues of $\Omega$.\\

Thus, energy spectrum of lowest LLs for double and triple-WSMs are $-\hbar k_z v_z +\frac{\mathcal{E}}{B}\hbar k_y \pm \lambda$ and $-\hbar k_z v_z+\frac{\mathcal{E}}{B}\hbar k_y$, $-\hbar k_z v_z +\frac{\mathcal{E}}{B}\hbar k_y \pm \sqrt{3}\lambda$ respectively.  Therefore, the n-fold degeneracy of lowest LLs is lifted due to the in-plane electric field $\mathcal{E}$ . The Eqns. (\ref{eqn_dwsms}),
(\ref{eqn_twsms}) and (34) are the modified energy spectrum of LLs and are our main results of the paper. The Landau levels spectrum for single, double and triple-WSMs with and without in-plane electric field are shown in Fig(\ref{figswsm}), Fig.(\ref{figdwsm}) and Fig.(\ref{figtwsm}) respectively. We have shown the Landau levels spectrum only near left chiral Weyl point. A similar LL spectrum can be easily obtained for right chiral Weyl point. It is clearly seen from  Fig.(\ref{figswsm} (b)), Fig.(\ref{figdwsm}(b)) and Fig.(\ref{figtwsm}(b)) that lowest Landau levels splits in the presence of $\mathcal{E}$ while it is unaffected without $\mathcal{E}$, see Fig.(\ref{figswsm} (a)), Fig.(\ref{figdwsm}(a)) and Fig.(\ref{figtwsm}(a)). However, there is always n number of chiral lowest LLs cut the zero energy which can be substantiated by the fact that in zero magnetic fields the topological charge of the Weyl nodes is $\pm n $ (the degeneracy of the zero LLs). Thus, the lowest LLs in double and triple WSMs cuts the zero energy along momentum $k_z$ at $\pm \lambda/{\hbar v_z}$ and 0, $\pm \sqrt{3}\lambda/{\hbar v_z}$ respectively. This LLs splitting modifies the density of states(DOS) which could be observed in quantum oscillations experiments. Next, we discuss the DOS of the multi-WSMs with and without electric fields and their physical consequences.

\section{Density of states}
\label{sec:4}
The density of states(DOS) governs the pattern of quantum oscillation measurements (through Shubnikov-de Haas effect or De Haas-van Alphen effect) of all transport and thermodynamic quantities \cite{lia16,pippard89,behnia11}.  In the presence of high magnetic fields, the energy dispersion of multi-WSMs in the plane perpendicular to $\textbf{B}$ is completely suppressed while it is still dispersive along the direction of the applied magnetic field. Thus the magnetic field step down the dimension of the system and the WSM in the external magnetic field can be visualized as a collection of one-dimensional systems with multiple subbands due to Landau levels. The DOS of multi-WSMs can be worked out by the following definition,

\begin{equation}
D(\varepsilon) = \frac{1}{2\pi l_B^2}\sum_{m}\int_{-\infty}^{+\infty} \frac{dk_z}{2 \pi}\,\delta[E_{m}^0(k_z)-\varepsilon], 
\label{dos_defn}
\end{equation}

\noindent where $ m $ labels Landau level index, $k_z$ is the conserved momentum of the one-dimensional conducting channel along the $\textbf{B}$ direction and $l_B$ is the magnetic length.\\

The DOS for multi- WSM in the absence of  in-plane electric field $\mathcal{E}$ can be obtained analytically. Let us first consider the dispersion for a single WSM 

\begin{equation}
E_{m}^0(k_z) = \sqrt{\hbar^2 v_z^2 k_z^2 + m \omega^2 \alpha_1^2}, \label{Eq:WeylDis}
\end{equation}

\noindent In such a system, each m$\geq$1 LLs crosses the Fermi energy $\varepsilon$ twice at critical momentum $k_{zc} = \pm \sqrt{\varepsilon^2-m \omega^2 \alpha_1^2}/(\hbar v_z)$, whereas the $m=0$ LL only cuts the Fermi energy once, at $k_z = -\varepsilon/(\hbar v_z)$. Therefore, the DOS for this single WSM is given by 

\begin{equation}
D(\varepsilon) = D_0 \left[1+\sum_{m\geq1}\frac{2\varepsilon\,\Theta(\varepsilon-\sqrt{m}\omega \alpha_1)}{\sqrt{\varepsilon^2-m\omega^2 \alpha_1^2}}\right], \label{Eq:DOS_swsm}
\end{equation}
where $D_0 = 1/(4\pi^2 l_B^2\hbar v_z)$ and $\Theta(x)$ is the Heaviside Theta function. \\

Similarly, the DOS for double WSM and triple WSMs can be obtainded analytically by the above same arguments. The DOS for double-WSM is given by 

\begin{equation}
    D(\varepsilon) = D_0\left[2+\sum_{m\geq2}\frac{2\varepsilon\,\Theta(\varepsilon-\sqrt{m(m-1)}\omega^2 \alpha_2)}{\sqrt{\varepsilon^2-m(m-1)\omega^4 \alpha_2^2}}\right], \label{Eq:DOS_dwsm}
\end{equation}

\noindent and for triple-WSM 

\begin{equation}
    D(\varepsilon) = D_0\left[3+\sum_{m\geq 3}\frac{2\varepsilon\,\Theta\left[\varepsilon-\sqrt{m(m-1)(m-2)}\omega^3 \alpha_3\right]}{\sqrt{\varepsilon^2-m(m-1) (m-2)\omega^6 \alpha_3^2}}\right], \label{Eq:DOS_twsm}
\end{equation}

\noindent where the numbers 2 and 3 in Eqns.(\ref{Eq:DOS_dwsm}) and (\ref{Eq:DOS_twsm}) accounted for the two-fold and three-fold degeneracy of the lowest LLs respectively. \\

\begin{figure}[h]       
\fbox{\includegraphics[scale=.5]{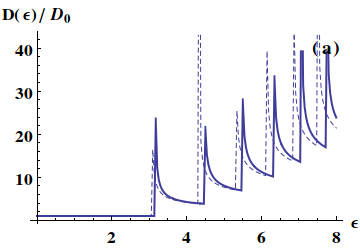}}   
\hspace{30px}
\fbox{\includegraphics[scale=.5]{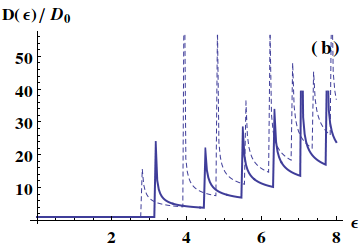}}
\hspace{30px}
\fbox{\includegraphics[scale=.5]{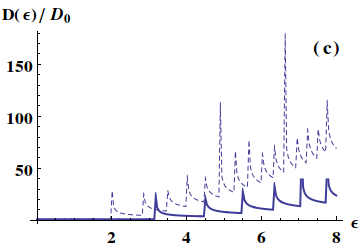}}
\caption{DOS for Landau levels in a single-WSM, renormalized by $D_0$ when (a) $\mathcal{E}=1$ (b) $\mathcal{E}=2$ and (c) $\mathcal{E}=3.5$ respectively. Dark and dotted plots shows the DOS without and with electric fields. Other parameters: B=5, $v_z$ =$1$, $\alpha_1=1$.}
\label{fig:dos_swsm}
\end{figure}

\begin{figure}[h]       
\fbox{\includegraphics[scale=.45]{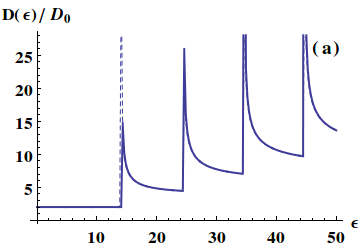}}   
\hspace{30px}
\fbox{\includegraphics[scale=.27]{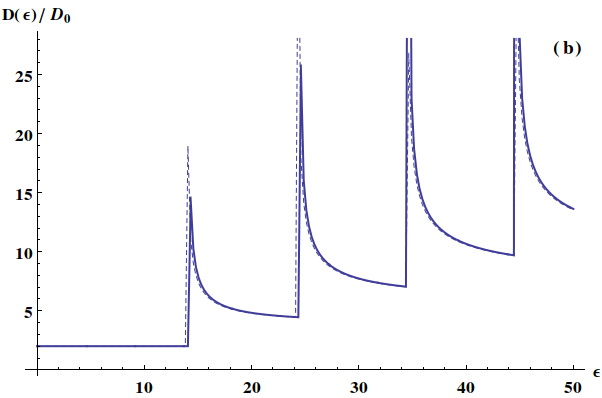}}
\hspace{30px}
\fbox{\includegraphics[scale=.27]{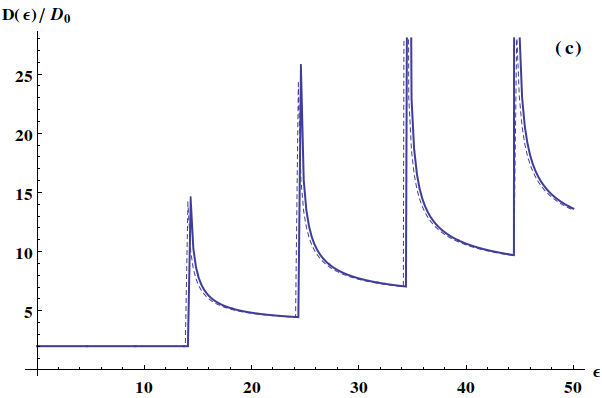}}
\caption{DOS for Landau levels in a double-WSM, renormalized by $D_0$ when (a) $\mathcal{E}=3$ (b) $\mathcal{E}=4$ and (c) $\mathcal{E}=5$ respectively. Dark and dotted plots shows the DOS without and with electric fields. Other parameters: B=5, $v_z$ =$1$, $\alpha_1=1$.}
\label{fig:dos_dwsm}
\end{figure}

\begin{figure}[h]       
\fbox{\includegraphics[scale=.25]{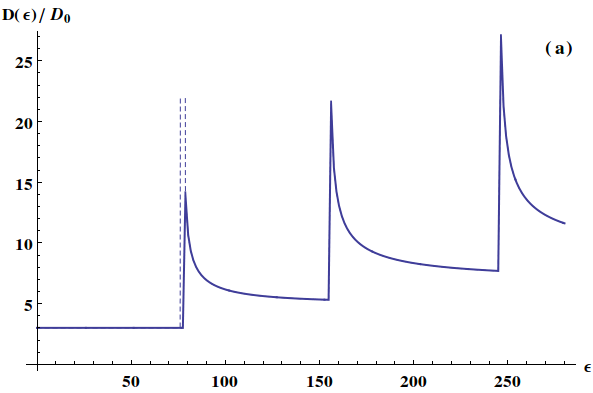}}   
\hspace{30px}
\fbox{\includegraphics[scale=.25]{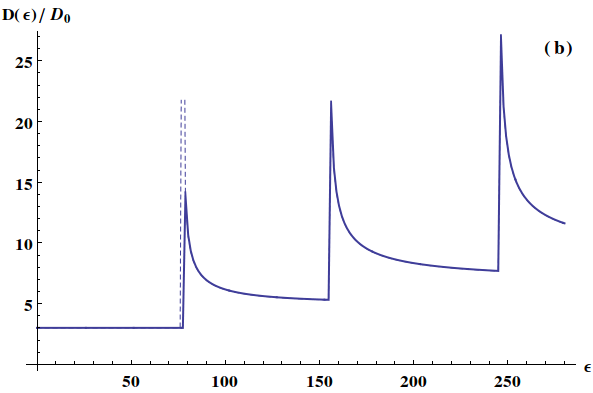}}
\hspace{30px}
\fbox{\includegraphics[scale=.25]{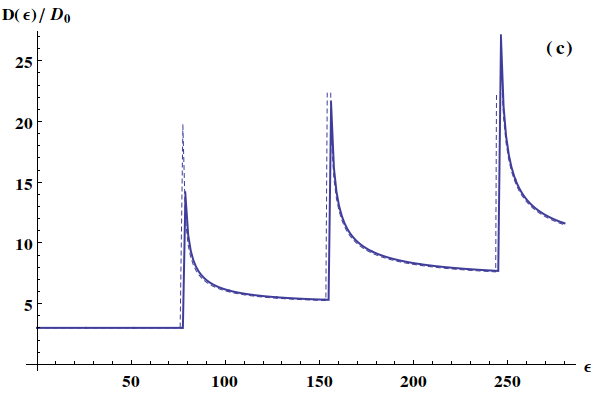}}
\caption{DOS for Landau levels in a triple-WSM, renormalized by $D_0$ when (a) $\mathcal{E}=5$ (b) $\mathcal{E}=10$ and (c) $\mathcal{E}=25$ respectively. Dark and dotted plots shows the DOS without and with electric fields. Other parameters: B=5, $v_z$ =$1$, $\alpha_1=1$.}
\label{fig:dos_twsm}
\end{figure}

In the presence of an in-plane electric field $\mathcal{E}$ along the x-direction, the DOS cannot be calculated analytically. Therefore, we compute it numerically using Eq.(\ref{dos_defn}) and display in Fig.(\ref{fig:dos_swsm}), Fig.(\ref{fig:dos_dwsm}) and  Fig.(\ref{fig:dos_twsm}) with increasing values of the electric field for single, double and triple-WSMs respectively. The magnetic oscillations have a sawtooth appearance originating from the $k_z$ dispersion of higher Landau levels (i.e. $m \geq n$). The peaks correspond to the spacing of the singularities which go like $B^(n/2)_{eff}$ where $B_{eff}$ is the effective or reduced effective magnetic field due to the electric field $\mathcal{E}$.  In the case of single-WSMs, $B_{eff}$ is reduced strongly compared to the bare applied magnetic field $B$ along z-direction whereas it has a minor modifications for the case of double and triple-WSMs. As a result, a considerable shift of peaks towards low $ \varepsilon$ are observed with increasing electric field $\mathcal{E}$ in the case for single-WSMs whereas there is a minor change in shift of peaks towards low $ \varepsilon$ are observed for the double and triple-WSMs case. In the case of single-WSM, the LLs becomes closer and closer with increasing $\mathcal{E}$  and at the critical value $\mathcal{E}_c=v_F\mathrm{B}$, it collapses in the plane perpendicular direction to B while it is still dispersive along B. The corresponding DOS squeezes with the electric field and at a critical value  $\mathcal{E}_c$, it reaches a constant value which corresponds to DOS of the one-dimension dispersion along the z-direction. For double and triple-WSMs, there are no such modifications in their DOS with $\mathcal{E}$ due to the non-collapse of the LLs and therefore, their DOS show small changes with $\mathcal{E}$. Further for lowest LLs (i.e. $m<n$), we observe that there is no change of the plateau in DOS at low energy $\epsilon$  due to the lifting of the degeneracy of their lowest LLs in the case of double and triple-WSMs. Since the degeneracy of lowest Landau level in the presence of in-plane electric field are symmetrically shifted about its lowest Landau levels energy $-\hbar v_z k_z$ in case of double and triple-WSMs. Therefore, when we add contributions from these shifted lowest Landau levels, it density of states(DOS) remains constants. These changes of DOS of multi-WSMs with $\mathcal{E}$ could be detected in angle-resolved quantum oscillations. e.g. the above features of DOS could be reflected in the specific heat and magnetization.

\section{Conclusion}
\label{sec:5}
In conclusion, we summarize the main findings of the present manuscript. We have analyzed a perturbative study of a multi-WSMs in crossed  electric and magnetic fields in low electric field approximation. This problem cannot be exactly solved for monopole charge $n> 1$ due to the absence of a reference frame in which the electric field vanishes. Therefore, we have calculated energy corrections up to second order. The main consequences of this electric field $\mathcal{E}$ are the n-fold degeneracy of the lowest Landau levels is lifted while the higher one remain unaffected due to the first order correction in electric field $\mathcal{E}$. The higher Landau levels have corrections due to the second order perturbation in electric field $\mathcal{E}$ while lowest Landau levels remain unaffected. We have compared the density of states of multi-WSMs system for both the absence and presence of the electric field. The lowest LLs (i.e. m$<$n)have no change of plateau in DOS at low energy $\epsilon$ in the case of double and triple-WSMs even with the lifting of the degeneracy of lowest LLs with electric field $\mathcal{E}$. Further, the higher LLs( m$ \geq$ n) contribute to the considerable amount of shift of peaks towards low energy $\epsilon$ with increasing electric field $\mathcal{E}$ for the case of a single-WSM while it has a little shift of peaks for double and triple WSMs case. This changes in DOS can be directly tested in angle-resolved quantum oscillation measurements in the future.\\

\section{Acknowledgements}

We acknowledge helpful discussions with Debanand Sa, N. S. Vidhyadhiraja and Daniel Yumnam. This work is supported by Science and Engineering Research Board (SERB), India for the SERB National Post doctoral Fellowship. I would like to thank JNCASR for the facilities for this work.\\

\section*{References}

\bibliography{mybibfile}

\end{document}